\documentclass{article}
\usepackage[margin=1.5 in]{geometry}
\usepackage{amsthm,amsmath}
\usepackage{amssymb}
\usepackage{graphicx}

\usepackage[T1]{fontenc}
\usepackage[utf8]{inputenc}
\usepackage{authblk} 
\title{Invariance of Spooky Action at a Distance in Quantum Entanglement under Lorentz Transformation}

\author{Mohammad Sharifi\footnote{Email:behsharifi@ut.ac.ir}}

\affil{Department of Physics, University of Tehran, Iran}

\begin{document}
\maketitle

\begin{abstract} 
We study the mechanism by which the particle-antiparticle entangled state collapses instantaneously at a distance. By making two key assumptions, we are able to show not only that instantaneous collapse of a wave function at a distance is possible but also that it is an invariant quantity under Lorentz transformation and compatible with relativity. In addition, we will be able to detect in which situation a many-body entangled system exhibits the maximum collapse speed among its entangled particles. Finally we suggest that every force in nature acts via entanglement.
\end{abstract}

\section{Introduction}

Assume that we have a particle-antiparticle pair in space, e.g., an electron and a positron that are entangled with one another. The electron is in location $(-x)$  and the positron in location $(+x)$ . Suppose that the two particles are stationary with respect to one another and to the observer's reference frame. Now, let us measure the spin of the electron at time $t_e$ , where  $t_e$ is measured with respect to our (the observer's) reference frame, and we find that it is down. One minute later, we measure the spin of the positron, and of course, we find that it is up. The question is at which time $t_p$  with respect to the observer's reference frame the wave function of the positron collapsed and the spin of the positron was set to spin up.

Because we measured the spin of the electron at time $t_e$, the electron is located at position $-x$ and the positron at $+x$ , and the velocity of each is zero, we expect from symmetry that $t_e$  should be equal to $t_p$. (Remember that $t_p$ is the time at which the wave function of the positron collapsed in the observer's reference frame). Thus, the speed of transfer of action from the electron to the positron should be 
\begin{equation}\label{1a} 
v=\frac{2x}{t_{e}-t_{e}}=\infty  
\end{equation} 
therefore, this speed is greater than the speed of light \cite{speed1,speed2,speed3}. However, what physical quantity has this speed?

\newtheorem{theorem}{Theorem}
\begin{theorem}\label{theorem1}
{The speed of action at a distance in quantum entanglement is the speed of the phase velocity of the entangled system.}
\end{theorem}

As we know, by using the de Broglie and Planck-Einstein relations and by considering the solution to the Klein-Gordon equation,
\begin{equation}\label{1j} 
\psi =\psi_{0}\exp i(\stackrel{\rightarrow}{k}.r-\omega{t}) 
\end{equation} 
the phase velocity of the wave function is determined by 
\begin{equation}\label{1m} 
v_{phase}=\frac{\omega}{k}=(\frac{E_{relativistic}}{\hbar}) / (\frac{P_{relativistic}}{\hbar})
\end{equation}
Generally, any solution to the wave equation of a free particle, whether a fermion or a boson has the form of equation \eqref{1j} (solution to Klein-Gordon equation) with different $\psi_{0}$.

\section{Unlocality}

Now, we want to determine the phase velocity of the electron and positron, but before we continue, we should consider a very important assumption.

\theoremstyle{definition}
\newtheorem*{conjecture}{Conjecture}
\begin{conjecture}\label{conjecture}
{The particle and anti-particle in quantum entanglement are not two entities but one entity.}
\end{conjecture}

In other words, we have assumed that the electron and positron have a shared energy and momentum. Although this assumption works, the above conjecture is not correct, especially if the two particles are not identical. In fact,the particle and antiparticle have an entangled wave function, not a common one. We will present a safer expression of this concept in theorem \ref{theorem2}. The merit of this assumption is its philosophical implication. By making this assumption, we have rejected the assumption of locality. If we suppose that the electron and positron are two particles, and we consider their phase velocities separately with no operation that considers some combination of the two velocities, our future equations will be Lorentz covariant.

By this method, we want to derive the speed of the phase velocity for the two particles. We suppose that the electron and positron are stationary in the observer's reference frame, and we assume that both are one particle; thus, we should add their energy and their momentum. 
\begin{equation}\label{2a} 
E=E_{1}+E_{2}=\frac{mc^{2}}{\sqrt{1-\beta_{1}^{2}}}+\frac{mc^{2}}{\sqrt{1-\beta_{2}^{2}}}=2mc^{2}  
\end{equation} 
\begin{equation}\label{2b} 
P=P_{1}+P_{2}=0 
\end{equation}
\begin{equation}\label{2c} 
v_{phase}=\frac{2mc^{2}}{0}=\infty  
\end{equation} 
which is equation \eqref{1a}. Note that $v_{1}$  and $v_{2}$  are zero. Recall that we assumed that the spin of the electron was measured at time $t_{e}$ ; recall also that we concluded that the wave function of the positron collapsed at the same time $t_{e}$  because the electron and positron were stationary and located, respectively, at positions $(-x)$ and $(+x)$  and because the phase velocity (in this case only) was infinite.

Now, we consider the proper time $t_{1}^\prime$ of the measurement of the spin of the electron in the reference frame of the electron and the proper time $t_{2}^\prime$ of the collapse of the positron spin wave function in the reference frame of the positron. Because the electron and positron are stationary with respect to the observer reference frame and with respect to each other, we obtain 
\begin{equation}\label{2d} 
t_{1}^\prime=t_{2}^\prime=t_{e}  
\end{equation}

\section{Lorentz Invariance of The Phase Velocity in Space-Time Coordinates} \label{Lorentz Invariance}

At this stage, we want to determine whether the instantaneous collapse at a distance, carried at the phase velocity, is invariant under Lorentz transformations. To prove this invariance, we consider a general case in which a pi meson moving with speed $v$  relative to the observer's reference frame decays at time $t_{m}$  and location $x_{m}$  into an electron, which moves with speed $v_{1}$  with respect to the observer's reference frame, and a positron, which moves with speed $v_{2}$  with respect to the observer's reference frame. Note that $v_{1}\neq v_{2}$  and that $v_{1}$ and $v_{2}$ may be either in opposite directions or in the same direction with respect to the observer's reference frame.

If the instantaneous collapse, which is mediated at the phase velocity,is invariant under Lorentz transformation, then the proper time  $t_{2}^\prime$ (the time in the reference frame of the positron when the positron receives the spooky action at the phase velocity) should depend only on the proper time $t_{1}^\prime$  (the time in the reference frame of the electron when the electron sends the action at the phase velocity), regardless of $v_{1}$ and $v_{2}$ or $(-v)$ , the speed of the observer's reference frame with respect to the pi meson.

Without loss of generality, we assume that after the decay of the pi meson, the electron moves with speed  $v_{1}$ to the left, the positron moves with speed  $v_{2}$ to the right,$v_{1}<v_{2}$. First we derive the phase velocity:
\begin{equation}\label{3a} 
\gamma_{1}=\frac{1}{\sqrt{1-\beta _{1}^{2}}}  
\end{equation} 
\begin{equation}\label{3b} 
\gamma_{2}=\frac{1}{\sqrt{1-\beta _{2}^{2}}}  
\end{equation} 
\begin{eqnarray}\label{3c} 
v_{phase}&=&\frac{E}{P}=\frac{E_{2}+E_{1}}{P_{2}+P_{1}}=\frac{mc^{2}\gamma_{2}+mc^{2}\gamma_{1}}{mv_{2}\gamma_{2}-mv_{1}\gamma _{1}}\nonumber\\
&=&\frac{(mc^{2}\gamma_{2}+mc^{2}\gamma_{1})}{(mv_{2}\gamma_{2}-mv_{1}\gamma_{1})}\frac{\sqrt{1-\beta_{2}^{2}}\sqrt{1-\beta_{1}^{2}}}{\sqrt{1-\beta_{2}^{2}}\sqrt{1-\beta_{1}^{2}}}\nonumber\\ 
&=&c^{2}\frac{\sqrt{1-\beta_{1}^{2}}+\sqrt{1-\beta _{2}^{2}}}{v_{2}\sqrt{1-\beta_{1}^{2}}-v_{1}\sqrt{1-\beta _{2}^{2}}}  
\end{eqnarray}

Because the electron is traveling to the left, we use $(-v_1)$ instead of $(v_1)$  in equation \eqref{3c}. We measure the spin of the electron at time $t_{1}$ , when the electron is in position $x_{1}$  ($t_{1}$, $x_{1}$,$t_{2}$ and $x_{2}$  are measured in the observer's reference frame);the electron then sends an action at the phase velocity of the system to the positron. The positron receives the action at time $t_{2}$ , when it is located at $x_{2}$. Notice that $t_1\neq{t_2}$ , and when the electron is located at $x_{1}$, the positron is located at $x_{b}$, where $x_{b}<x_{2}$ . The action travels at the phase velocity, which is not always equal to infinity, and while the action travels from $x_{1}$  to $x_{2}$ , the positron travels from $x_{b}$ to $x_{2}$.

Now, we want to prove that the proper time $t_{1}^\prime$  in the reference frame of the electron only depends on the proper time $t_{2}^\prime$  in the reference frame of the positron. Using Lorentz transformations, we can write 
\begin{equation}\label{3d} 
t_{1}=t_{m}+\frac{t_{1}^\prime-(v_{1}x_{1}^\prime/c^{2})}{\sqrt{1-\beta _{1}^{2}}}=t_{m}+\frac{t_{1}^\prime}{\sqrt{1-\beta_{1}^{2}}} 
\end{equation} 
\begin{equation}\label{3e} 
t_{2}=t_{m}+\frac{t_{2}^\prime+(v_{2}x_{2}^\prime/c^{2})}{\sqrt{1-\beta_{2}^{2}}} =t_{m}+\frac{t_{2}^\prime}{\sqrt{1-\beta_{2}^{2}}} \end{equation} 
\begin{equation}\label{3f} 
x_{1}=x_{m}+\frac{x_{1}^\prime-v_{1}t_{1}^\prime}{\sqrt{1-\beta_{1}^{2}}}=x_{m}+\frac{-v_{1}t_{1}^\prime}{\sqrt{1-\beta_{1}^{2}}} 
\end{equation}
and 
\begin{equation}\label{3g} 
x_{2}=x_{m}+\frac{x_{2}^\prime+v_{2} t_{2}^\prime}{\sqrt{1-\beta_{2}^{2}}}=x_{m}+\frac{v_{2}t_{2}^\prime}{\sqrt{1-\beta_{2}^{2}}}  
\end{equation} 

Note that the position of the electron in its reference frame $(x_{1}^\prime)$  and that of the positron in its reference frame $(x_{2}^{\prime})$  are zero. The distance that the action needs to travel is 
\begin{eqnarray}\label{3h}
\Delta{x}&=&x_{2}-x_{1}=\frac{v_{2}t_{2}^\prime}{\sqrt{1-\beta_{2}^{2}}}-\frac{-v_{1}t_{1}^\prime}{\sqrt{1-\beta_{1}^{2}}}\nonumber\\
&=&\frac{v_{2}t_{2}^\prime\sqrt{1-\beta_{1}^{2}}+v_{1}t_{1}^\prime\sqrt{1-\beta_{2}^{2}}}{\sqrt{1-\beta_{1}^{2}}\sqrt{1-\beta_{2}^{2}}}  
\end{eqnarray} 
In addition, the time it takes for the action to travel from $x_{1}$  to $x_{2}$  is 
\begin{eqnarray}\label{3i}
\Delta{t}&=&t_{2}-t_{1}=\frac{t_{2}^\prime}{\sqrt{1-\beta_{2}^{2}}}-\frac{t_{1}^\prime}{\sqrt{1-\beta_{1}^{2}}}\nonumber\\ 
&=&\frac{t_{2}^\prime\sqrt{1-\beta_{1}^{2}}-t_{1}^\prime\sqrt{1-\beta_{2}^{2}}}{\sqrt{1-\beta_{1}^{2}}\sqrt{1-\beta_{2}^{2}}} 
\end{eqnarray}
If we suppose that the action is mediated at the phase velocity, we can write 
\begin{equation}\label{3j} 
\Delta x=\Delta{t}.v_{phase}  
\end{equation}
By substituting equations \eqref{3c}, \eqref{3h}, \eqref{3i}  into equation \eqref{3j} and after some calculations we obtain 
\begin{equation}\label{3q}
(t_{2}^{\prime}-t_{1}^{\prime})[v_{2}^{2}c^{2}-v_{1}^{2} v_{2}^{2}-c^{4}+c^{2} v_{1}^{2}-(c^{2}+v_{1}v_{2})(\sqrt{c^{2}-v_{1}^{2}} \sqrt{c^{2}-v_{2}^{2}})]=0 
\end{equation} 
Because the second term on the left-hand side of equation \eqref{3q} is not zero, we conclude that although $t_{1}\ne t_{2}$ ,
\begin{equation}\label{3r} 
t_{1}^{\prime} =t_{2}^{\prime}  
\end{equation} 
which is a very important result (please look at \eqref{2d}); thus, we can state the following: 
\newtheorem{corollary}{Corollary} 
\begin{corollary}\label{corollary1}
{In the collapse at a distance of an entangled identical particle-antiparticle pair, which occurs at the phase velocity, the proper time of the collapse of one particle's wave function in its own reference frame is invariant under Lorentz transformation and is equal to the proper time of the measurement of the state of the other antiparticle in its own reference frame.}
\end{corollary}
Because the particle and antiparticle are similar to each other, we expected by symmetry that the proper time of sending the action in one particle's reference frame and the proper time of receiving the action in the other particle's reference frame should be equal. Thus, this result can be observed as a vindication of theorem \ref{theorem1} and the following theorem \ref{theorem2}.

\section{Phase Velocity of The Entangled Systems In Space-Time Coordinates}

\begin{theorem}\label{theorem2}
{If one specification of the wave function (spin, color, isospin, etc.) of a many-body system is entangled among several particles, for the calculation of the phase velocity of the collapse of that specification in space-time coordinates, it is safe to assume that all entangled particles are one entity with one total energy and momentum with respect to that specification.}
\end{theorem}

This behavior is similar to that of two entangled fermions participating in superconductivity, which behave statistically as one bosonic entity and obey Bose-Einstein condensation \cite{cooper1,cooper2}. The phase and group velocities of a particle are parameters of its wave function. To calculate the phase velocity of the entangled particles, we must first construct their entangled wave function. The entangled system is a type of many-body problem. There is a general solution to construct the wave function of a many-body system. For identical fermions, we use the Slater determinant, and thus, the final wave function will be antisymmetric. For example, for two identical fermions, we write
\begin{equation}\label{4a} 
\psi=\frac{1}{\sqrt{2}}
\begin{bmatrix}
\psi_{a}(r_{a})&\psi_{a}(r_{b})\\
\psi_{b}(r_{a})&\psi_{b}(r_{b})
\end{bmatrix}
=\psi_{a}(r_{a})\psi_{b}(r_{b})-\psi_{a}(r_{b})\psi_{b}(r_{a}) 
\end{equation} 
Because the wave functions of the particles $(a)$  and $(b)$  have the form (consider Klein-Gordon solution)
\begin{equation}\label{4b}  
\psi_{n}(r_{m})=\psi_{0n}(r_m)\exp i(k_{n}.r-\omega_{n}t) 
\end{equation} 

where  $(n,m=1,2)$ , equation \eqref{4b} can be written as 
\begin{equation}\label{4c} 
[\psi_{0a}(r_a)\psi_{0b}(r_b)-\psi_{0a}(r_b)\psi_{0b}(r_a)]\exp i(k_{a}.r-\omega_{a}t)\exp i(k_{b}.r-\omega_{b}t) 
\end{equation} 
Here,  $\psi_{0a}$ and $\psi_{0b}$  are matrices and $k_{a}$  and  $k_{b}$ are vectors. Equation \eqref{4c} has a more compact form:
\begin{equation}\label{4d} 
\psi=\psi_{0}\exp i[(k_{a}+k_{b}).r-(\omega_{a}+\omega_{b})t] 
\end{equation} 
Thus, the phase velocity of $\psi$ is 
\begin{equation}\label{4e}  
v_{phase}=\frac{\omega_{a}+\omega_{b}}{k_{a}+k_{b}}  
\end{equation} 

We used $r$  instead of $r_{m}$  in the exponential term of equation  \eqref{4b}. If we want to be more rigorous, we can define two variables $R$  and $\Delta R$ :
  
\begin{equation} \label{4f} 
R=\frac{r_{a}+r_{b}}{2}  
\end{equation} 
\begin{equation} \label{4g} 
\Delta R=\frac{r_{a}-r_{b}}{2}  
\end{equation} 
Equation \eqref{4c} then becomes
\begin{eqnarray} \label{4h} 
&&[\psi_{0a}(r_a)\psi_{0b}(r_b)\exp i(k_{a}-k_b).\Delta R-\psi_{0a}(r_{b})\psi_{0b}(r_{a})\exp -i(k_{a}-k_{b}).\Delta R]\nonumber\\
&&\times\exp i[(k_{a}+k_b).R-(\omega_{a}+\omega_b)t]  
\end{eqnarray}

Although $R$  is the coordinate of space, $\Delta R$  is only a vector and does not correspond to any point in space, so the phase velocity is determined only by the second line of equation \eqref{4h} and equation \eqref{4e}.

For a many-body system composed of identical bosons, the net wave function is symmetric. For example, for two identical bosons, instead of equation \eqref{4a}, we write 
\begin{equation} \label{4i} 
\psi=\psi_{a}(r_a)\psi_{b}(r_b)+\psi_{a}(r_b)\psi_{b}(r_a) 
\end{equation}

which is the \emph{Permanent of matrix}\cite{permanent} (symmetric Slater determinant with addition instead of subtraction of terms). Again, however, the phase velocity of the system is equation \eqref{4e}, because the exponential terms in equation \eqref{4d} do not change, and the only part of the wave function that changes is $\psi_{0}$.

For non-identical particles, we simply multiply wave functions and do not care about symmetrization (Hartree product\cite{hartree}). Both the permanent and the Slater determinant are composed of the addition or subtraction of these Hartree products. In general, if the system is composed of both fermions and bosons (non-identical or identical), the general form of the wave function can have mixed symmetric and antisymmetric parts or only a single term, which describes a system with completely non-identical particles. The net wave function is again the addition or subtraction of Hartree product terms. For example, for a three-body system,

\begin{equation}\label{4j} 
\psi=\psi_{i}(r_l)\psi_{j}(r_m)\psi_{k}(r_n)\pm...\quad (i,j,k,l,m,n=1,2,3) 
\end{equation} 
Because each individual particle's wave function should take part one time in every Hartree product term in the series, we conclude from equation\eqref{4b} that each Hartree product term has an identical exponential expression. This means that the series $\psi$ has the form 
\begin{equation}\label{4k} 
\psi=[\psi_{0i}(r_l)\psi_{0j}(r_m)\psi_{0k}(r_n)\pm...]\exp i[(k_{1}+k_{2}+k_{3}).r-(\omega_{1}+\omega_{2}+\omega_{3})t] 
\end{equation} 
so the phase velocity of the net wave function can be written as 
\begin{equation}\label{4l} 
v_{phase}=\frac{\omega}{\stackrel{\rightarrow}{|k|}}=\frac{\sum_{i=1}^{n}\omega_{i}}{|\sum_{i=1}^{n}\stackrel{\rightarrow}{k}_{i}|}  
\end{equation} 
regardless of whether the particles in the many-body system are identical or non-identical or have Fermi-Dirac or Bose-Einstein statistics. Thus we can modify our false conjecture, prove theorem \ref{theorem2} and write 
\begin{equation}\label{4m} 
v_{phase}=\frac{E}{\stackrel{\rightarrow}{|P|}}=\frac{\sum_{i=1}^{n}E_{i}}{|\sum_{i=1}^{n}\stackrel{\rightarrow}{P}_i|}  
\end{equation}

Now, suppose that in the future we create a quantum entanglement system that is composed of three or more particles with different masses entangled simultaneously with each other; then theorems \ref{theorem1} and \ref{theorem2} are applicable to this system, but instead of corollary \ref{corollary1}, we have another corollary.
\begin{corollary}\label{corollary2}
{Suppose there is a fictional quantum system composed of several particles with different masses where each particle is moving with an arbitrary velocity in an arbitrary direction, and the system is entangled in one specification of the wave function; then, only for an observer in the reference frame in which the summation of the momenta of all of the entangled particles is zero, the speed of sending and receiving action among particles is infinite.}
\end{corollary}

Corollary \ref{corollary2} applies to every particle, even particles with zero mass. there is one interesting related point: Suppose that we have three particles $a$, $b$  and $c$; particles $a$  and $b$  are entangled in $x$  specification of the net wave function; and particles  $a$ and  $c$ are entangled in $y$  specification of the net wave function. The energy and momentum of particle $c$  do not play any part in the phase velocity of the collapse of the $x$ specification of the wave function, and those of particle $b$  have no effect on the phase velocity of the collapse of the $y$ specification of the wave function.

\section{Phase Velocity in Energy-Momentum Spaces}

What we observed in section \ref{Lorentz Invariance} was this: We considered a system in which two particles, say an electron and a positron, were far from one another but had a localized shared wave function $\psi(x,t)$ . Part of $\psi$  was located at $(x_e,t_e)$, and the other part was at $(x_p,t_p)$. The two parts of $\psi$  were interacting with each other at space-time coordinate $(x,t)$  at the phase velocity in this space. Now, there should not be anything special about $(x,t)$ . In other words, although we live in  $(x,t)$ coordinates, and $x$  and $t$  are allowed to vary in our macroscopic world, a quantum wave equation does not discriminate between an $(x,t)$  coordinate and any other type of coordinate, similar to energy-momentum, and thus, the result should be valid in other coordinates. As a result, similar versions of theorems \ref{theorem1} and \ref{theorem2} should be valid in all coordinates. 
  
Now, suppose that we are in energy-momentum coordinates and that the wave function is $\phi(p,E)$ . Part of the wave function is located at $(p_e,E_e)$  and the other part at $(p_p,E_p)$ , but $\phi$ has a specific phase velocity in energy-momentum space, which is determined from the $(x)$  and  $(t)$ of each part of $\phi(p,E)$ . In other words, the electron-positron pair has a shared $(x,t)$  in energy-momentum space because we considered them as one identity in this space by the shared wave function $\phi(p,E)$ (theorem \ref{theorem2}). What we are doing is not completely correct. In the non-relativistic Schrodinger Wave Equation, the $\hat{x}$  can be considered as an operator, but $\hat{t}$  is not an operator. The probability that $\psi$  exists in a specific time is meaningless and always equal to one. In the relativistic wave equation, both $x$  and $t$  are parameters, and the $\hat{x}$  no longer can act as an operator.

Because the electron and positron are identical particles with the same mass, To determine the phase velocity in energy-momentum space, we should add the $(x)$ and $(t)$  in the exponential term of $\phi(p,E)$  and then divide by each other. Because in momentum space, we have 
\begin{equation}\label{5a}
i\hbar\frac{\partial\phi(p)}{\partial p}=x \phi(p)
\end{equation}
the above equation contains the solution. 
\begin{equation}\label{5b}
\phi(p)=\phi_0(p)\exp(\frac{-i}{\hbar}x.p)
\end{equation}
As before,to construct the entangled wave function, we again multiply both $\phi_1$  and $\phi_2$ :
\begin{eqnarray}\label{5c-1}
\phi(p)&=&\phi_{01}(p)\exp(\frac{-i}{\hbar}x_{1}.p)\phi_{02}(p)\exp(\frac{-i}{\hbar}x_{2}.p)\nonumber\\
&=&\phi_1(p)\phi_2(p)=\phi_0(p)\exp\frac{-i}{\hbar}[(x_1+x_2).p]\
\end{eqnarray}
We can create a similar incorrect equation in energy space,
\begin{equation}\label{5c-2}
\phi(E)=\phi_0(E)\exp\frac{i}{\hbar}[(t_1+t_2).E]
\end{equation}
and then define the phase velocity in energy-momentum space. As a result, we obtain
\begin{equation}\label{5c-3}
\phi(p,E)=\phi_0(p,E)\exp\frac{-i}{\hbar}[(x_1+x_2).p-(t_1+t_2).E]
\end{equation}

From equations \eqref{3d} to \eqref{3g} we have 
\begin{eqnarray}\label{5d}
x_1+x_2&=&x_{m}+\frac{-v_{1}t_{1}^\prime}{\sqrt{1-\beta_{1}^{2}}}+x_{m}+\frac{v_{2}t_{2}^\prime}{\sqrt{1-\beta_{2}^{2}}}\nonumber\\
&=&2x_m+\frac{v_{2}t_{2}^\prime\sqrt{1-\beta_{1}^{2}}-v_{1}t_{1}^\prime\sqrt{1-\beta_{2}^{2}}}{\sqrt{1-\beta_{1}^{2}}\sqrt{1-\beta_{2}^{2}}} 
\end{eqnarray} 
and
\begin{eqnarray}\label{5e}
t_1+t_2&=&t_{m}+\frac{t_{1}^\prime}{\sqrt{1-\beta_{1}^{2}}}+ t_{m}+\frac{t_{2}^\prime}{\sqrt{1-\beta_{2}^{2}}}\nonumber\\ 
&=&2t_m+\frac{t_{2}^\prime\sqrt{1-\beta_{1}^{2}}+t_{1}^\prime\sqrt{1-\beta_{2}^{2}}}{\sqrt{1-\beta_{1}^{2}}\sqrt{1-\beta_{2}^{2}}} 
\end{eqnarray}
If we assume that there is a correspondence between the $(x,t)$ and $(p,E)$  coordinates, we have 
\begin{equation}\label{5f}
(ict,x,y,z)\longrightarrow(\frac{E}{ict},p_x,p_y,p_z)
\end{equation}
Now, the problem of considering $x$  and $t$  as operators appears. To avoid the problem, we suppose that $x_m=t_m=0$ (the space-time coordinate of the disintegration of the pi meson and the creation of the particle pair should be considered the central point of the Cartesian coordinate system) and obtain (look at \eqref{5c-3})
\begin{equation}\label{5g}
v_{phase(p,E)}=\frac{t_1+t_2}{x_1+x_2}=\frac{t_{2}^\prime\sqrt{1-\beta_{1}^{2}}+t_{1}^\prime\sqrt{1-\beta_{2}^{2}}}{v_{2}t_{2}^\prime\sqrt{1-\beta_{1}^{2}}-v_{1}t_{1}^\prime\sqrt{1-\beta_{2}^{2}}}
\end{equation}
The distance that the spooky action must travel in energy-momentum space is 
\begin{equation}\label{5h}
\Delta{P}=v_{phase(p,E)}.{\Delta{E}}
\end{equation}
Replacing $p_1$, $p_2$, $E_1$  and $E_2$  by looking at \eqref{3c} and multiplying the denominators with numerators, we obtain 
\begin{eqnarray}\label{5i}
\frac{\Delta{P}}{\Delta{E}}&=&\frac{P_{2}-P_{1}}{E_{2}-E_{1}}=\frac{mv_{2}\gamma_{2}+mv_{1}\gamma _{1}}{mc^{2}\gamma_{2}-mc^{2}\gamma_{1}}\nonumber\\
&=&\frac{(mv_{2}\gamma_{2}+mv_{1}\gamma_{1})}{(mc^{2}\gamma_{2}-mc^{2}\gamma_{1})}\frac{\sqrt{1-\beta_{2}^{2}}\sqrt{1-\beta_{1}^{2}}}{\sqrt{1-\beta_{2}^{2}}\sqrt{1-\beta_{1}^{2}}}\nonumber\\ 
&=&\frac{v_{2}\sqrt{1-\beta_{1}^{2}}+v_{1}\sqrt{1-\beta _{2}^{2}}}{c^{2}(\sqrt{1-\beta_{1}^{2}}-\sqrt{1-\beta _{2}^{2}})}
\end{eqnarray}
Combining equations \eqref{5g}, \eqref{5h} and \eqref{5i}, and after doing a little algebra finally we can derive  
\begin{eqnarray}\label{5l}
&&t_{2}^\prime[(c^{2}-v_2^2)(c^2-v_{1}^{2})-({c^{2}+v_1v_2})\sqrt{1-\beta_{1}^{2}}\sqrt{1-\beta_{2}^{2}}]\nonumber\\
&=&t_{1}^\prime[(c^{2}-v_1^2)(c^2-v_{2}^{2})-(c^{2}+v_1v_2)\sqrt{1-\beta_{2}^{2}}\sqrt{1-\beta_{1}^{2}}]
\end{eqnarray}
which again gives \eqref{2d} and \eqref{3r} $t_{1}^\prime=t_{2}^\prime$. Thus, the action can travel at the phase velocity in energy-momentum space, and there is nothing special about space-time coordinates.
From \eqref{3j} and \eqref{5h}, we obtain
\begin{equation}\label{5m}
\frac{x_2-x_1}{t_2-t_1}=\frac{E_2+E_1}{p_2+p_1} \quad \& \quad \frac{t_2+t_1}{x_2+x_1}=\frac{p_2-p_1}{E_2-E_1}
\end{equation}
Note that the first equation is not merely the reciprocal of the second one. Some elements have changed sign, and the validity of \eqref{3j} does not imply the validity of \eqref{5h}. If, in equations \eqref{3j} and \eqref{5h}, we substitute $cp$  and  $ct$ instead of $p$  and $t$ , it can be shown that for entangled system composed of \emph{subluminal} particles the following is always true:
\begin{equation}\label{5n}
v_{phase(p,E)}>{1}  \quad \& \quad    v_{phase(x,t)}>{1}
\end{equation}
From a classical point of view, in energy-momentum space, the positron does not change its energy and momentum and is frozen when the action travels from the electron to the positron, so the speed of action $v_{phase(p,E)}$ need not be bigger than unity to exceed the positron in energy momentum space. from the classical point of view, for fixed energy and momentum $E$ and $p$ , the particle can increase its $t$  and $x$ , but the reverse is not true (for fixed $x$ and $t$ the particle can not increase its energy momentum). on the other hand in quantum theory, the energy-momentum is undetermined or variable until the wave function $\phi_{p,E}$ collapse. this is the reason that the phase velocity in energy momentum space $v_{phase(p,E)}$ for subluminal particle should be bigger than unity \eqref{5n}. the action should exceed particles even in energy momentum space.

Suppose that both particles are moving to the right direction with a superluminal speed. It can be shown that for this entangled system composed of superluminal particles the phase velocity is less than unity in energy momentum space and is subluminal in space time coordinate. There is a question that should be answered; is this a necessary condition for the collapse of entangled wave function that the phase velocity of the system $v_{phase(x,t)}$ be bigger than its group velocity? Why the phase velocity of time like particles $v_{phase(x,t)}$, should be in the uncausality region of space time and should be superluminal. Can the entangled wave function of the above-described system collapse? How can the electron which is located on the left informs the superluminal positron which is located on the right. how the subluminal action exceeds the superluminal positron? Bear in mind that the direction of phase velocity is still from left to right. It seems that, the superluminal electron positron pair no longer can make an entangled wave function. what about a single independent particle wave function. Why we have not yet observed a single superluminal particle?

\section{Entanglement as Virtual Particles and Mediators}
\begin{figure}
	\centering
		\includegraphics[width=3in]{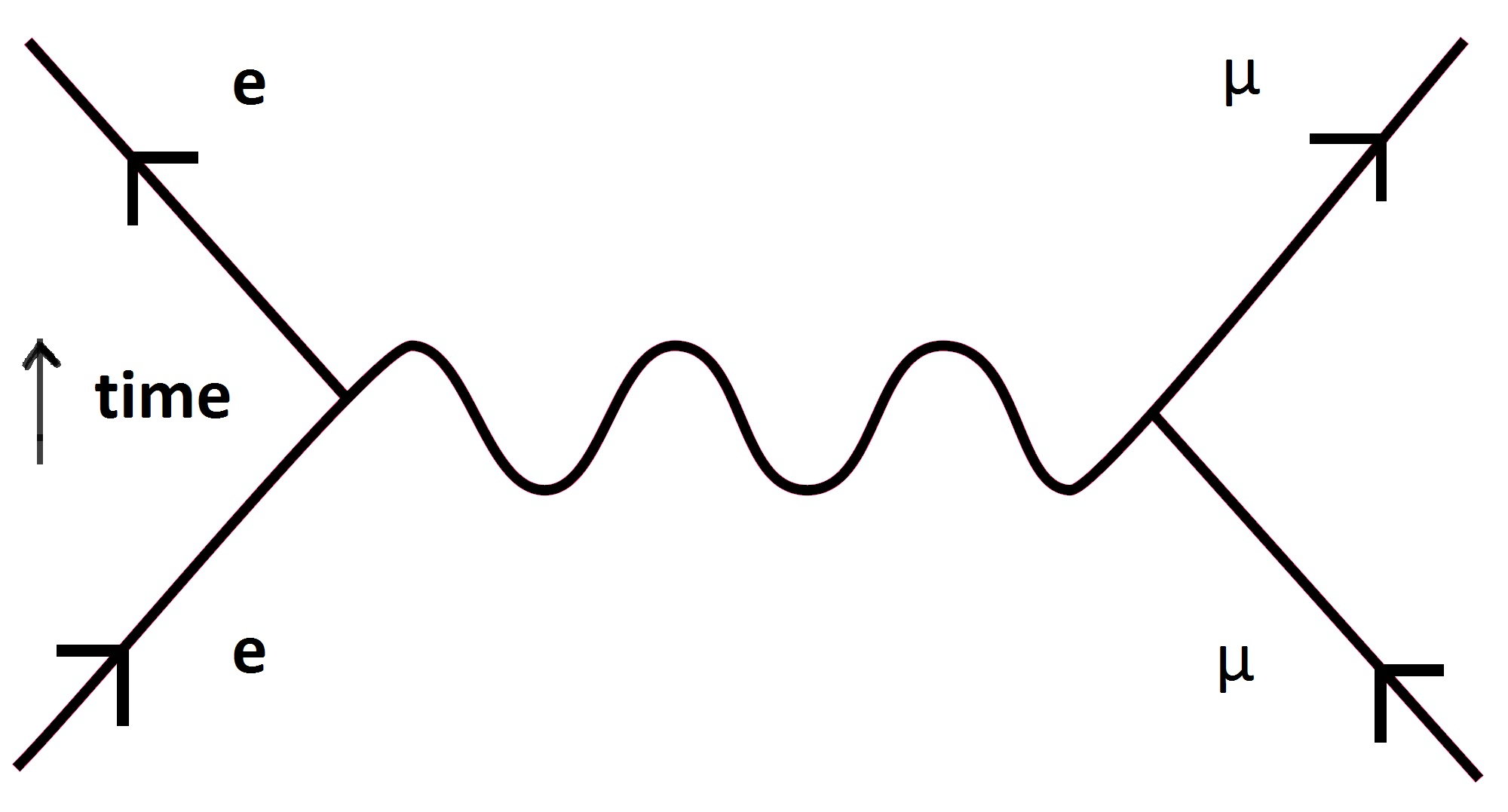}
	\label{Feynman.jpg}
	 \caption{Feynman diagram of electron muon scattering}
\end{figure}
\begin{theorem}\label{theorem3}
{The speed of exchanged virtual particle between real particles is the speed of the phase velocity of the entangled system.}
\end{theorem}

In this section, we will address the subject of the interaction force between particles. When two particles interact,they may only gain or lose momentum and energy; their masses remain constant. As a result, the mediator should have real momentum and energy but imaginary mass. Thus, accepted mediator theory states that mediators should be space-like and superluminal or off-mass shell. As we know the mediator cannot transfer information.

Consider two similar particles, e.g., an electron and muon in Feynman diagram, with the following energy-momenta:
\begin{equation}\label{7b} 
P_{e}=(E_{e},ip_{e}) 
\end{equation} 
\begin{equation}\label{7c} 
P_{e}^\prime=(E_{e}^\prime,ip_{e}^\prime) 
\end{equation} 
\begin{equation}\label{7d} 
P_{\mu}=(E_{\mu},ip_{\mu}) 
\end{equation} 
and
\begin{equation}\label{7e} 
P_{\mu}^\prime=(E_{\mu}^\prime,ip_{\mu}^\prime) 
\end{equation} 
Where $P_{e}$  and $P_{\mu}$  are the energy-momenta of the electron and muon before the exchange of the virtual photon, respectively, and $P_{e}^\prime$  and $P_{\mu}^\prime$  are the energy-momenta after the exchange of the virtual photon, respectively. The electron is located on the left and the muon on the right far from each other, and both particles are moving to the right. in addition $e$ and $e'$ intersect at the left vertex and $\mu$ and $\mu'$ intersect at the right vertex of the Feynman diagram.

From energy-momentum conservation in Feynman diagram, we obtain 
\begin{equation}\label{7f} 
P_{e}+P_{\mu}=P_{e}^\prime+P_{\mu}^\prime   \qquad        P=(E,ip) 
\end{equation} 
The energy-momentum of the virtual photon is  $p_q$, and from conservation laws,
\begin{equation}\label{7g}
P_{e}=P_{q}+P_{e}^\prime   \quad \longrightarrow    \quad      P_{q}=P_{e}-P_{e}^\prime=P_{\mu}^\prime-P_{\mu}   \qquad        P=(E,ip) 
\end{equation} 
where we have assumed that the electron sends the virtual photon, and the muon receives it. The direction of movement of the virtual particle is from left to right. The energy and momentum of the virtual photon are
\begin{equation}\label{7h} 
E_q=\frac{m_{e}c^{2}}{\sqrt{1-\beta_{e}^{2}}}-\frac{m_{e}c^{2}}{\sqrt{1-\beta_{e}^{\prime 2}}}  
\end{equation} 
and
\begin{equation}\label{7i} 
p_{q}=\frac{m_{e}v_{e}}{\sqrt{1-\beta_{e}^{2}}}-\frac{m_{e}v_{e}^\prime}{\sqrt{1-\beta _{e}^\prime}}  
\end{equation} 
so its speed is 
\begin{equation}\label{7j} 
v_{virtual}=c^{2}\frac{p_q}{E_q}=c^{2}(\frac{m_{e}v_{e}}{\sqrt{1-\beta_{e}^{2}}}-\frac{m_{e}v_{e}^\prime}{\sqrt{1-\beta_{e}^\prime}})/(\frac{m_{e}c^{2}}{\sqrt{1-\beta_{e}^{2}}}-\frac{m_{e}c^{2}}{\sqrt{1-\beta _{e}^{\prime 2}}}) 
\end{equation}
In addition, its mass $m_q$  can be calculated as
\begin{equation}\label{7k} 
m_{q}^{2}=(P_{e}-P_{e}^\prime)^{2} =(E_{e}-E_{e}^\prime)^{2}-(p_{e}-p_{e}^\prime)^{2}  
\end{equation} 
which $m_{q}$  can easily be proven to be imaginary \cite{ryder}.

Consider the speed $v_1$  defined by 
\begin{equation}\label{7l}
v_1=\frac{E_{e}^\prime+E_{e}}{p_{e}^\prime+p_{e}}=(\frac{m_{e}c^{2}}{\sqrt{1-\beta_{e}^{\prime 2}}}+\frac{m_{e}c^{2}}{\sqrt{1-\beta_{e}^{2}}})/(\frac{m_{e}v_{e}^{\prime}}{\sqrt{1-\beta_{e}^{\prime 2}}}+\frac{m_{e}v_{e}}{\sqrt{1-\beta_{e}^{2}}}) 
\end{equation} 

We want to show that  $v_1$ \eqref{7l}  is equal to the speed of the virtual particle emitted by the electron \eqref{7j}. Assuming that the two equations are equal, we are led to 
\begin{eqnarray}\label{7m}
&&c^{2}(\frac{m_{e}v_{e}}{\sqrt{1-\beta_{e}^{2}}}-\frac{m_{e}v_{e}^\prime}{\sqrt{1-\beta_{e}^{\prime2}}})(\frac{m_{e}v_{e}^\prime }{\sqrt{1-\beta_{e}^{\prime2}}}+\frac{m_{e}v_{e}}{\sqrt{1-\beta_{e}^{2}}})\nonumber\\ 
&&=(\frac{m_{e}c^{2}}{\sqrt{1-\beta_{e}^{2}}}-\frac{m_{e}c^{2}}{\sqrt{1-\beta_{e}^{\prime2}}})(\frac{m_{e}c^{2}}{\sqrt{1-\beta _{e}^{\prime2}}}+\frac{m_{e}c^{2}}{\sqrt{1-\beta_{e}^{2}}}) 
\end{eqnarray} 
By expanding the above equation and doing some mathematical calculations, we obtain 
\begin{eqnarray}\label{7p}
&&\frac{m_{e}^{2}c^{2}(1-\beta_{e}^{\prime2})}{1-\beta_{e}^{\prime2}}+\frac{m_{e}^2(c^{2}-v_{e}^{\prime}v_{e})}{\sqrt{1-\beta_{e}^{\prime2}}\sqrt{1-\beta_{e}^{2}}}\nonumber\\
&&=\frac{m_{e}^{2}(c^{2}-v_{e}^{\prime}v_{e})}{\sqrt{1-\beta_{e}^{\prime2}}\sqrt{1-\beta_{e}^{2}}}+\frac{m_{e}^2 c^{2}(1-\beta_{e}^{2})}{1-\beta_{e}^{2}}\end{eqnarray} 
which proves the validity of equation \eqref{7m}. Thus, we can write 
\begin{equation}\label{7q}
v_1=\frac{E_{e}^\prime+E_{e}}{p_{e}^\prime+p_{e}}=c^{2}\frac{p_{e}^\prime-p_{e}}{E_{e}^\prime-E_{e}}
\end{equation}
In a similar manner, we can prove 
\begin{equation}\label{7r}
v_2=\frac{E_{\mu}^\prime+E_{\mu}}{p_{\mu}^\prime+p_{\mu}}=c^{2}\frac{p_{\mu}^\prime-p_{\mu}}{E_{\mu}^\prime-E_{\mu}}
\end{equation}

Using \eqref{7g} in \eqref{7q} and \eqref{7r}, we find that the right-hand sides of \eqref{7q} and \eqref{7r} are the same, $v_1=v_2$, so we can algebraically add the numerators and denominators of the left-hand sides and obtain
\begin{equation}\label{7s}
\frac{E_{e}^\prime+E_{e}+E_{\mu}^\prime+E_{\mu}}{p_{e}^\prime+p_{e}+p_{\mu}^\prime+p_{\mu}}=c^{2}\frac{p_{e}^\prime-p_{e}}{E_{e}^\prime-E_{e}}=c^{2}\frac{p_{\mu}^\prime-p_{\mu}}{E_{\mu}^\prime-E_{\mu}}=v_1=v_2
\end{equation}
Finally by using \eqref{7f} and \eqref{7g} in the above equation, we obtain
\begin{equation}\label{7t}
v_{phase}=\frac{2(E_{e}+E_{\mu})}{2(p_{e}+p_{\mu})}=c^{2}\frac{p_{e}^\prime-p_{e}}{E_{e}^\prime-E_{e}}=c^{2}\frac{p_{q}}{E_{q}}=v_{virtual}
\end{equation}

Thus, the speed of all mediators is the phase velocity of the system, and as a result, all four principal interaction forces act via quantum entanglement. The result we have obtained here is logical. Two field particles cannot interact unless they form a shared wave function, which results from sharing one specification of the entangled wave function (for example, the electron-muon shared wave function via entanglement in the phase of the electric charge due to the gauge invariance of the wave function).

In equation \eqref{7t}, note that if the signs of both the $p_q$ and $E_q$  of the virtual particle are reversed, the phase velocity does not change direction. Both particles have sent it and received it simultaneously. The signs of the momentum and even energy of the virtual particle can vary, depending on the speed of the reference frame relative to the speed of particles because it is a superluminal effect. In other words, in a specific reference frame the electron has sent the virtual particle, and the muon has received it but in another reference frame, the electron is a receiver, and muon is the sender. The final word about virtual particle is that although the energy momentum of virtual particle is completely undetermined, its velocity and trajectory are completely determinable.

There should be a similarity between superluminal quantum tunneling \cite{tunneling2,virtual,tunneling3} or all other superluminal quantum phenomena and exchange of virtual particle between real particles or collapse of wave function in quantum entanglement. The next equation \eqref{6ze} suggests that the phase velocity of the system is the only invariant superluminal velocity in space time coordinate that behave simillar to $c$.

\section{Discussions and Conclusions} \label{discussions}

The EPR paradox \cite{epr} helped us to obtain profound insight about nonlocality in quantum mechanics in the past decades. There is a well-established theorem \cite{bell} that rejects the principle of locality in quantum phenomena. In addition after Bell inequality several experiments have been done in favor of nonlocality
in quantum mechanics and against hidden local variable theories \cite{nonlocality experiment}. However, does these experiments and theorem reject the validity of special theory of relativity and permit exchange of  information at superluminal speeds? On the other hand, there is an unfriendly behavior between general theory of relativity and quantum mechanics. In this section, we discuss about the consequences that can be driven from the results that we obtained about the wave function in the past sections.

As we previously saw the phase velocity elements $(\omega ,k)$  in \eqref{3j} were obtained from the de Broglie and Planck-Einstein relations in quantum theory \eqref{1m}, but $\Delta x$  and  $\Delta t$ in \eqref{3d} to \eqref{3i} were obtained from Lorentz transformation and relativity. Thus, quantum mechanics and relativity cooperate exactly show here that the entangled system can be compatible with both concepts. This cooperation was the core of our entire derivation. 

Now let's discuss about the possibility of existence of an independent superluminal particle. However, first we should discuss about the un causality of entanglement. In our entangled system in section \ref{Lorentz Invariance}, suppose that the electron is located on the left, and positron is located on the right. In addition, again suppose that both particles are moving to the right direction. Thus, the direction of the phase velocity of the entangled system is from left to right too, how can the measurement of the spin of the positron, which is located on the right, lead to the collapse of the wave function of the electron, which is located on the left? In fact, if we are in such a reference frame, we will observe that before we measure the spin of the positron at time $t_{p}$ on the right, the electron wave function has already collapsed at time $t_{e}$ on the left, where  $t_{e}<t_{p}$. Thus, the action travels with the phase velocity, from left to right. In other words, the electron was aware of what we planned to do with positron in the future, and it collapsed beforehand.

The answer to this dilemma is that although there is cause and effect in every physical interaction, the particle and antiparticle do not constitute cause and effect in the case of entanglement. In other words, you cannot say that the collapse of the positron wave function is due to the collapse of the electron wave function or vice versa. Both of them are cause, and both of them are effect. You are dealing with one particle (entity), not two. The concepts of locality and cause-and-effect are valid for objects slower than the speed of light. In the above example, you can change the speed of your reference frame and be in a reference frame that both particles are moving to the left then you observe that the measurement of the spin of the positron wave function has led to the collapse of the electron wave function with the correct directionality of the phase velocity, from right to left, which seems logical.

There is a peculiar characteristic of the phase velocity that behaves exactly like $c$. Suppose that we have a source with speed $u$  and that it radiates some wave at a speed of $f(u)$. If $f(u)$ is truly the speed of the radiation, the equation below should be valid:
\begin{equation}\label{6ze} 
f(\frac{u+v}{1+uv/c^2})=\frac{f(u)+v}{1+f(u)v/c^2}
\end{equation}

However, if we use $f(x)=2x$, equation \eqref{6ze} does not work. The above equation is valid only for three functions $f(x)$:  $f(x)=x$, $f(x)=c$  and $f(x)=\frac{c^2}{x}$. The first, $f(x)=x$, does not lead to a new equation. We conclude that $f(x)=c$ and $f(x)=\frac{c^2}{x}$  have the same characteristic. Thus, $v_{phase}$ truly acts as a radiation speed. As we can see, the phase velocity plays the same rule in quantum mechanics as the speed of light plays in relativity and the group velocity plays in classical mechanics. 

Note that although we proposed that the superluminal action could be responsible for the collapse of all members of entangled systems, we cannot conclude that it can be possible for us to send a truly single particle or any kind of informative entity, faster than the speed of light to transfer an information at a superluminal speed from one location in space to another location.

When the $\psi_{space}$ of a single unentangled wave function collapse, it communicates by its phase velocity to other locations in space-time that the wave function should not collapse in other locations (recall the copenhagen interpretation of quantum mechanics), and because the phase velocity is superluminal, there is no causality, and we cannot perform the trick of changing the observer's reference frame to make it possible for the wave function to collapse at two different locations. That is why it is impossible to detect the wave function in the second location (for example, a far galaxy) after the collapse of wave function in the first location.  Note that this internal communication is done at infinite velocity in the reference frame of collapsed wave function.

Theoretically, the collapse of the wave function of a single superluminal particle is impossible because its phase velocity of collapse is \emph{subluminal} and obeys causality. There is a big question about the detection of a superluminal particle by a detector\cite{opera}. how we can collapse the wave function $\psi_{space}$ of a superluminal particle without making it possible for the particle to communicate at once (at infinite velocity in the reference frame of particle) to all locations of the universe to tell that its wave function cannot be collapsed in any other  locations (space like regions). If the phase velocity of particle is in the time-like region of space-time, then the collapse of $\psi_{space}$ of particle on the earth does not make any restriction for a detector to find the particle after a second in another galaxy or another space-like region of space time. The phase velocity should be superluminal. It is a necessary condition for the collapse of wave function.

Thus no informative mass and energy can exceed  ${\rm c}$. If we define the action as something that contains information, the action cannot be superluminal. The wave function state is not an informative energy or mass. Any change in the wave function (collapse) does not mean a transfer of informative mass, energy or any type of information, and the use of the term \emph{action} to indicate a phenomenon that involves a transfer of information is false. To avoid the movement of a cloud of particles at the phase velocity due to a collapse, it is better to interpret $\psi\psi_{space}^*$ as a probability density rather than a density of a cloud of particles or a matter density.

You can never travel faster than a superluminal particle, but you can easily change the speed of the reference frame in such a way that the superluminal velocity approaches $+\infty$. However, if you reach that point and continue, the speed of the particle jumps from $+\infty$  to $-\infty$. Thus, when a particle is created at location $A$ traveling faster than light and is then detected in location $B$, you can easily change the reference frame in such a way that it seems that the particle is first detected by the detector at $B$  and then travels to the left and is created at  $A$, which is meaningless. This is possible and meaningful only if locations  $A$ and $B$ are one location in energy-momentum space, and the detector and creator are one subject and one entity in space-time coordinates. This scenario is only possible for entangled actions and virtual particles, which are related to both sources. No truly independent particle or information can ever travel at this speed from a creator to a detector. This particle will be ubiquitous. In other words, there are only three speeds in nature: first, zero, related to all subluminal particles; second, $c$; and third, infinity, related to all superluminal effects. When there is a superluminal-independent particle in a reference frame moving from left to right, it is in fact a ghost and a spook.

Finally, we discuss about the general theory of relativity. As we saw only non-informative virtual particles can move faster than light. The virtual particle is truly a virtual subject and is sent neither from electron to positron nor from positron to electron. However, any entity either real or virtual should interact with the gravitational field. This is because the gravitational field  concerns the curvature in the medium that through which other masses, energies, information and non informative actions travel. Because every movement in a distorted medium has a distorted trajectory, every physical quantity that has either real mass such as an independent particle or imaginary mass such as a virtual particle or even a non informative exchanged action  in quantum entanglement with no stress energy tensor, logically should interacts with gravitational field. In addition as we previously knew the gravitational field could affect the phase of the wave function \cite{gravitation}. It seems that the gravitational field can affect all entities. (recall that matter waves are not physical entities). it also seems that
the quantization of gravitational field is vain. why we have not yet detected any massless spin 2 boson?

In contrast, due to this geometrical trajectory, the study of the action and the collapse of wave functions between a part of a system inside a Schwarzschild radius \cite{hobson,misner} and another part of it outside the radius should be interesting. How can a mediator that has a real trajectory at the phase velocity in a metric escape this metric and affects other objects outside this radius? can particles tunnel outside the event horizon (recall equation \eqref{6ze}). The collapse of wave function is not a metaphysical subject; its action should to \emph{travel in space-time at the phase velocity of the system.}

\bibliographystyle{plain}

\end{document}